\documentclass[12pt]{article}
\usepackage{amssymb,amsmath,amsthm,amsfonts,amscd}
 \usepackage{graphicx}
\newcommand\be{\begin{equation}}
\newcommand\ee{\end{equation}}
\newcommand\bea{\begin{eqnarray}}
\newcommand\eea{\end{eqnarray}}

\newcommand{\fatalpha}{{\bf \alpha \kern -0.44em \alpha}}
\newcommand{\fatsigma}{{\bf \sigma \kern -0.54em \sigma}}
\newcommand{\tpchi}{{\bf D \kern -0.35em D}}
\newcommand{\llambda}{{\bf \lambda \kern -0.45em \lambda}}

\renewcommand{\theequation}{\arabic{equation}}
\renewcommand{\theequation}{\thesection.\arabic{equation}}
\bibliography{plain}
\title{\bf \large{Investigating a Class of $2\otimes2\otimes d$ Chessboard
Density Matrices via Linear and Non-linear Entanglement Witnesses
Constructed by Exact Convex Optimization}} \vspace{20mm}
\author{ M. A. Jafarizadeh $^{a,b,c}$
 \thanks{E-mail:jafarizadeh@tabrizu.ac.ir}  ,
 Y. Akbari  $^{a,b}$
 \thanks{E-mail:y-akbari@tabrizu.ac.ir} ,
  K. Aghayar  $^{a}$
 \thanks{E-mail:aghayar@tabrizu.ac.ir},\\
  A. Heshmati  $^{a}$
 \thanks{E-mail:heshmati@tabrizu.ac.ir},
  M. Mahdian  $^{a}$
 \thanks{E-mail:mahdian@tabrizu.ac.ir},
\\ $^a${\small Department of Theoretical Physics and Astrophysics,
Tabriz University, Tabriz 51664, Iran.} \\ $^b${\small Institute for
Studies in Theoretical Physics and Mathematics, Tehran 19395-1795,
Iran.} \\ $^c${\small Research Institute for Fundamental Sciences,
Tabriz 51664, Iran. }} \pagebreak

\pagebreak

\vspace{20mm}

\begin{document}
\maketitle \vspace{15mm}
\newpage


\begin{abstract}
Here we consider a class of $2\otimes2\otimes d$ chessboard density
matrices starting with three-qubit ones which have positive partial
transposes with respect to all subsystems. To investigate the
entanglement of these density matrices, we use the entanglement
witness approach. For constructing entanglement witnesses (EWs)
detecting these density matrices, we attempt to convert the problem
to an exact convex optimization problem. To this aim, we map the
convex set of separable states into a convex region, named feasible
region, and consider cases that the exact geometrical shape of
feasible region can be obtained. In this way, various linear and
non-linear EWs are constructed. The optimality and decomposability
of some of introduced EWs are also considered. Furthermore, the
detection of the density matrices by introduced EWs are discussed
analytically and numerically.


{\bf PACs Index: 03.65.Ud }

\end{abstract}

\newpage

\section{Introduction}
Bound entangled states, states with positive partial transposes with
respect to all subsystems, are of great importance in quantum
information processes \cite{horod1,wang1,horod3,horod4,wolf1}. One
class of bound entangled states is the three-qubit chessboard states
considered in \cite{acin} where the authors have used a separability
criterion due to P. Horodecki to show the boundness of such states.
The boundness of these states for some range of parameters are also
investigated in \cite{hyllus1} using entanglement witnesses (EWs)
and in \cite{eisert} from the perspective of convex optimization.
Another class of chessboard states has been discussed in
\cite{pitt1} again by using entanglement witnesses (EWs). The EWs
are of special interest since it has been proved that for any
entangled state there exists at least one EW detecting it. The EWs
are Hermitian operators which have non-negative expectation values
over all separable states while they have negative expectation
values over, that is they are able to detect, some entangled states
\cite{woron1,horod2}.
\par
In this paper, we consider a generalized form of the above
chessboard states initially for $2\otimes2\otimes2$  case, then
extend them for $2\otimes2\otimes d$ case and use EWs approach to
analyze their entanglement. For constructing the relevant EWs, we
attempt to convert the problem to an exact convex optimization
problem. This method are general and one can apply it for
multi-qubits in a similar way. All of witnesses constructing in this
way are valid with some changes in notation. As the dimension of
problem increases the number and categories of EW's increases but
the procedures are same in general. Convex optimization techniques
have been widely used in quantum information problems recently
\cite{moor1,moor2,brand1,doherty2,doherty3,ja6,doherty4,shor1,ja7,ja8,ja9,ja10}.
In references \cite{ja1,ja2,ja3,ja4,ja5} the problem of constructing
EWs was converted to a linear programming problem, a special case of
convex optimization problem, exactly or approximately. To this aim,
the convex set of separable states was mapped into a convex region,
named feasible region (FR). The FR may be a polygon by itself or it
may not. When FR was not a polygon, it was approximated by a
polygon. In this way, the problem was converted to a linear
programming problem whose linear constraints came from the exact or
approximated boundary surfaces of FR.
\par
Here we consider the cases that the geometrical shape of FR can be
obtained exactly and hence convert the problem to an exact convex
optimization problem. Any hyper-plane tangent to the FR
corresponds to a linear EW. According to the geometrical shape of
FR, we can construct non-linear EWs or can not. It is shown that
when the geometrical shape of FR is a polygon, all EWs are linear;
otherwise it is possible to construct non-linear EWs. In the
previous works where a non-polygonal FR was approximated by a
polygonal one, the number of obtained linear EWs was not
sufficient for constructing non-linear EWs. However, in the
present work where we consider the exact geometrical shape of a
non-polygonal FR, any hyper-plane tangent to the surface of FR is
a linear EW. Therefore, there exist innumerable linear EWs which
is enough for constructing a non-linear EW as the envelop of
linear functionals arising from them. By construction, a
non-linear EW plays the role of innumerable linear EWs as a whole
and hence it may detect bound entangled states. Our approach is
typical and can be applied in all cases where the exact
geometrical shape of FR is known.
\par
The paper is organized as follows. In Section 2, we review the basic
notions and definitions of EWs relevant to our study and describe
our approach  of constructing EWs. Then we present a generalized
form of a class of three-qubit density matrices of \cite{acin}. In
Section 3, we consider the construction of linear and non-linear EWs
that can detect the mentioned density matrices. Section 4 is devoted
to an analysis of optimality of introduced EWs. It is proved that
some of the EWs are optimal. In Section 5, we consider the detection
of mentioned density matrix by introduced EWs analytically and
numerically. Section 6 is devoted to the comparison of our results
with other works. In section 7 we extend all these methods to
$2\otimes2\otimes d$ case and we see that all these methods are
general and one can apply them for multipartite chessboard density
matrices. This extension neither change the structure of PPT's
conditions nor the EW's structures. In section 8 numerical analysis
for detection ability of introduced EW's for $2\otimes2\otimes2$ and
$2\otimes2\otimes3$ chessboard density matrices are discussed.


\section{Preliminaries}
\subsection{\small{A class of three-qubit density matrices with positive partial transposes}}
Here we consider a generalized form of a class of three-qubit
density matrices presented in \cite{acin}
\begin{equation}\label{ppt}
 \rho=\frac{1}{n} \left(%
 \begin{array}{cccccccc}
           a & 0 & 0 & 0 & 0 & 0 & 0 &r_{_{1}} e^{i \varphi_{_{1}}} \\
           0 & b & 0 & 0 & 0 & 0 &r_{_{2}} e^{i \varphi_{_{2}}} & 0 \\
           0 & 0 & c & 0 & 0 &r_{_{3}} e^{i \varphi_{_{3}}} & 0 & 0 \\
           0 & 0 & 0 & d &r_{_{4}} e^{i \varphi_{_{4}}} & 0 & 0 & 0 \\
           0 & 0 & 0 &r_{_{4}} e^{-i \varphi_{_{4}}} & \frac{1}{d} & 0 & 0 & 0 \\
           0 & 0 &r_{_{3}} e^{-i \varphi_{_{3}}} & 0 & 0 & \frac{1}{c} & 0 & 0 \\
           0 &r_{_{2}}e^{-i \varphi_{_{2}}} & 0 & 0 & 0 & 0 & \frac{1}{b} & 0 \\
           r_{_{1}} e^{-i \varphi_{_{1}}} & 0 & 0 & 0 & 0 & 0 & 0 & \frac{1}{a} \\
 \end{array}%
\right)
\end{equation}
where $a,b,c,d$ are non-negative parameters, $0\leq r_i \leq 1$
for $i=1,2,3,4$ and  $n=( \
a+b+c+d+\frac{1}{a}+\frac{1}{b}+\frac{1}{c}+\frac{1}{d} \ )$. It
is easy to see that this density matrix has positive partial
transposes with respect to all subsystems, i.e., it is a PPT
state. The density matrix of \cite{acin} is a special case of
$\rho$ where $\varphi_1=0, \ r_{_{1}}=1, \
r_{_{2}}=r_{_{3}}=r_{_{4}}=0$, and $a=1$. We want to show that for
some values of the parameters, $\rho$ is a PPT entangled state. To
this aim, we will construct various linear and non-linear
non-decomposable EWs that are able to detect it.
\par
Written in the Pauli matrices basis, $\rho$ has the form
\begin{equation}\label{dens}
\begin{array}{c}
  \rho=\frac{1}{8}[III+r_{_{300}}\sigma_{z}II+r_{_{030}}I\sigma_{z}I+
  r_{_{003}}II\sigma_{z}+r_{_{330}}\sigma_{z}\sigma_{z}I+r_{_{303}}\sigma_{z}I\sigma_{z}\\
  \hspace{1.2cm}+r_{_{033}}I\sigma_{z}\sigma_{z}+r_{_{333}}\sigma_{z}\sigma_{z}\sigma_{z}
    +r_{_{111}}\sigma_{x}\sigma_{x}\sigma_{x}+r_{_{112}}\sigma_{x}\sigma_{x}\sigma_{y}
  +r_{_{121}}\sigma_{x}\sigma_{y}\sigma_{x}\\
  \hspace{1.5cm}+r_{_{211}}\sigma_{y}\sigma_{x}\sigma_{x}+
  r_{_{122}}\sigma_{x}\sigma_{y}\sigma_{y}+r_{_{212}}\sigma_{y}\sigma_{x}\sigma_{y}+
  r_{_{221}}\sigma_{y}\sigma_{y}\sigma_{x}+r_{_{222}}\sigma_{y}\sigma_{y}\sigma_{y}] \\
\end{array}
     \end{equation}
where the coefficients $r_{_{ijk}}$ are given in appendix B. We
will try to construct our non-decomposable EWs by using Pauli
group operators appearing in the $\rho$. But before this, let us
review the basic notions and definitions of EWs relevant to our
study.

\subsection{\small{Entanglement witnesses}}
Let us first recall the definition of entanglement and
separability \cite{terhal1}. By definition, an n-partite quantum
mixed state $\rho\in {\cal{B}}({\cal{H}})$ (the Hilbert space of
bounded operators acting on the Hilbert space
$\mathcal{H}={\cal{H}}_{d_{1}}\otimes...\otimes{\cal{H}}_{d_{n}})$
is called fully separable if it can be written as a convex
combination of pure product states, that is
\begin{equation}\label{fullsep}
    \rho=\sum_{i} p_{i} | \alpha_{i}^{(1)} \rangle \langle \alpha_{i}^{(1)} |\otimes
    | \alpha_{i}^{(2)} \rangle \langle \alpha_{i}^{(2)}
    |\otimes...\otimes| \alpha_{i}^{(n)} \rangle \langle \alpha_{i}^{(n)} |
\end{equation}
where $|\alpha_{i}^{(j)}\rangle$ are  arbitrary but normalized
vectors lying in the  $\mathcal{H}_{d_{j}}$, and $p_{i}\geq0$ with
$\sum_{i}p_{i}=1$. Otherwise, $\rho$ is called entangled.
Throughout the paper, by separability  we mean fully separability.
\par
An entanglement witness (EW) W is a Hermitian operator which has
non-negative expectation value over all separable states
$\rho_{s}$ and its expectation value over, at least, one entangled
state $\rho_{e}$ is negative. The existence of an EW for any
entangled state is a direct consequence of Hahn-Banach theorem
\cite{rudin1} and the fact that the  subspace of separable density
operators is convex and closed.
\par
Based on the notion of partial transpose map, the EWs are
classified into two classes: decomposable (d-EW) and
non-decomposable (nd-EW). An EW W is called decomposable if there
exist positive operators $\mathcal{P}, \mathcal{Q}_{K}$ such that
\begin{equation}
 W=\mathcal{P}+\sum_{K\subset
 \mathcal{N}}\mathcal{Q}_{K}^{T_{K}}
\end{equation}
where $\mathcal{N}:=\{1,2,3,...,n\}$ and  $T_{K}$ denotes the
partial transpose with respect to  partite $K\subset \mathcal{N}$
and it is non-decomposable if it can not be written in this form
\cite{doherty1}. Clearly, d-EWs can not detect bound entangled
states (entangled states with positive partial transpose (PPT)
with respect to all subsystems) whereas there are some bound
entangled states which can be detected by an nd-EW.
\par
A non-linear EW associated to an entangled density matrix
$\varrho$ is simply a non-linear functional of $\varrho$ such that
it is non-negative valued over all separable states, but has
negative value over the density matrix $\varrho$. A non-linear EW
can be viewed as the envelop of a set of linear functionals
$Tr(W\varrho)$ that arise from corresponding linear EWs W.
\par
Usually one is interested in finding EWs W which detect entangled
states in an optimal way. An EW W is called an optimal EW if there
exists no other EW which detects more entangled states than W. It
is shown that the necessary and sufficient condition for
optimality of an EW W is that there exist no positive operator
$\mathcal{P}$ and $\epsilon> 0$ such that $W'=W-\epsilon
\mathcal{P}$ be again an EW \cite{lewen1}.
\subsection{\small{Constructing of EWs via exact convex optimization}}
Let us consider a set of given Hermitian operators $Q_{_{i}}$. By
using these operators, we will attempt to construct various linear
and non-linear EWs. To this aim, for any separable state
$\rho_{s}$ we introduce the maps
\begin{equation}\label{varp}
    P_{_{i}}=Tr(Q_{_{i}}\rho_{s})
\end{equation}
which map the convex set of separable states into a convex region
named the feasible region (FR). Any hyper-plane tangent to the FR
corresponds to a linear EW, since such hyper-planes separate the FR
from entangled states. Hence, we need to determine the geometrical
shape of FR. In general, determining the geometrical shape of FR is
a difficult task. However, one may choose the Hermitian operators
$Q_{_{i}}$ such that the exact geometrical shape of FR can be
obtained rather simply. By such a choice, when the FR is a polygon,
its surface corresponds to linear EWs which are linear combinations
of the operators $Q_{_{i}}$; otherwise, linear EWs come from any
hyper-plane tangent to the surface of FR. When the FR is not a
polygon, besides the linear EWs it is possible to obtain non-linear
EWs for the given density matrix.
\par
To obtaine the geometrical shape of FR, we note that every
separable mixed state $\rho_{s}$ can be written as a convex
combination of pure product states, so the subspace of separable
states $\mathcal{S}$ can be considered as a convex hull of the set
of all pure product states $\mathcal{D}$. Thus first we specify
the geometrical shape of a region obtained from mapping of
$\mathcal{D}$ under the $P_{i}$'s. If the resulted region is
convex by itself, we get the FR, otherwise we have to take the
convex hull of that region as FR.
\par
In this paper, the operators $Q_{_{i}}$ are chosen as linear
combinations of Hermitian operators in the Pauli group
$\mathcal{G}_{n}$, a group consisting of tensor products of the
identity $I_{2}$ and the usual Pauli matrices
$\sigma_{x},\sigma_{y}$ and $\sigma_{z}$ together with an overall
phase $\pm1$ or $\pm i$ \cite{preskill,gott1,gott2}.



\section{A class of three-qubit EWs}
In this section,we want to introduce various nd-EWs for the
density matrix $\rho$ of (\ref{ppt}). To simplify the analysis,
let us classify these EWs according to the shape of relevant FRs:
polygonal, conical, cylindrical and spherical. Hereafter, we will
use the following notation for the three-qubit Pauli group
operators
\begin{equation}\label{notat}
    O_{ijk}=\sigma_{i}\otimes\sigma_{j}\otimes\sigma_{k},\qquad
    i,j,k=0,1,2,3,
\end{equation}
where $\sigma_{0}$, $\sigma_{1}$, $\sigma_{2}$ and $\sigma_{3}$
stand for the $2\times2$ identity matrix $I_{2}$ and single qubit
Pauli matrices $\sigma_{x}$, $\sigma_{y}$, $\sigma_{z}$
respectively.
 Let us begin
with polygonal case.
\subsection{\small{EWs with polygonal FR}}
Let us consider the following operators
$$
Q_{_{1}}^{^{\mathrm{Po}}}=O_{333},\quad
Q_{_{2}}^{^{\mathrm{Po}}}=O_{111}+(-1)^{i}O_{122},\quad
Q_{_{3}}^{^{\mathrm{Po}}}=O_{212}+(-1)^{i+1}O_{221},\quad i=0,1
$$
and try to construct nd-EWs from them for detecting $\rho$. To
this end, we define the maps
$$
P_{_{j}}=Tr(Q_{_{j}}^{^{\mathrm{Po}}}|\alpha\rangle|\beta\rangle|\gamma\rangle\langle\alpha|\langle\beta|\langle\gamma|)
,\quad j=1,2,3
$$
for any pure product state
$|\alpha\rangle|\beta\rangle|\gamma\rangle$. In this case, the FR
is a polygon which its boundary planes are as follows:
\begin{equation}\label{hp}
    (-1)^{j_{_{1}}}P_{1}+(-1)^{j_{_{2}}}P_{2}+(-1)^{j_{_{3}}} P_{3}=1\quad,\quad
    (j_{_{1}},j_{_{2}},j_{_{3}})\in \{0,1\}^{3}
\end{equation}
(for a proof, see appendix A). These planes can be rewritten as
$$
\min_{_{|\alpha\rangle|\beta\rangle|\gamma\rangle}}Tr([III-(-1)^{j_{_{1}}}Q_{_{1}}^{^{\mathrm{Po}}}
-(-1)^{j_{_{2}}}Q_{_{2}}^{^{\mathrm{Po}}}-(-1)^{j_{_{3}}}Q_{_{3}}^{^{\mathrm{Po}}}]
|\alpha\rangle|\beta\rangle|\gamma\rangle\langle\alpha|\langle\beta|\langle\gamma|)=0
$$
It is seen that the operators in the bracket have non-negative
expectation values over all pure product states, hence they give
rise to the following linear EWs
\begin{equation}\label{lew}
    ^{1}W_{i_{_{1}}i_{_{2}}i_{_{3}}i_{_{4}}}^{^{\mathrm{Po}}} =III+(-1)^{i_{_{1}}}O_{333}+(-1)^{i_{_{2}}}O_{111}
    +(-1)^{i_{_{3}}}O_{122}+(-1)^{i_{_{4}}}O_{212}+(-1)^{i_{_{2}}+i_{_{3}}+i_{_{4}}+1}O_{221},
\end{equation}
where $(i_{_{1}},i_{_{2}},i_{_{3}},i_{_{4}})\in\{0,1\}^{4}$.
Besides the above 16 EWs, we can construct other 16 EWs by using
the fact that local unitary operators take an EW to another EW.
For this purpose, we act the phase-shift gate
$$
M= \left(
\begin{array}{cc}
  1 & 0 \\
  0 & i \\
\end{array}
\right)
$$
locally on the first qubit which takes $
\sigma_{_{x}}\longrightarrow\sigma_{_{y}},
\quad\sigma_{_{y}}\longrightarrow-\sigma_{_{x}}$, and
$\sigma_{_{z}}\longrightarrow\sigma_{_{z}}$ under conjugation, and
get
\begin{equation}\label{lew1}
    \begin{array}{c}
\hspace{-4cm}
^{2}W_{i_{_{1}}i_{_{2}}i_{_{3}}i_{_{4}}}^{^{\mathrm{Po}}}
    =MII(W_{i_{_{1}}i_{_{2}}i_{_{3}}i_{_{4}}}^{^{\mathrm{Po}}})M^{\dagger}II=III+(-1)^{i_{_{1}}}O_{333}\\
 \hspace{.8cm}+(-1)^{i_{_{2}}}O_{111}+(-1)^{i_{_{3}}}O_{122}
 +(-1)^{i_{_{4}}+1}O_{212}+(-1)^{i_{_{2}}+i_{_{3}}+i_{_{4}}}O_{221}.\\
    \end{array}
\end{equation}
We could replace $Q_{1}^{^{\mathrm{Po}}}$ with the operator
$\sigma_{z}\sigma_{z}I$ or any cyclic permutation of it, but since
these lead to  d-EWs we do not consider such cases here.
\par
In this way, we have constructed 32 linear EWs with polygonal FR.
\subsection{\small{EWs with conical FR}}
For this case, we consider the following Hermitian operators
$$
\hspace{-2cm}  Q_{_{1}}^{^{\mathrm{Co}}}=O_{k'j'l'},\quad
  Q_{_{2}}^{^{\mathrm{Co}}}=O_{111}+(-1)^{i}O_{kjl},\quad
Q_{_{3}}^{^{\mathrm{Co}}}=O_{lkj}+(-1)^{i}O_{jlk},\quad i=0,1,\\
$$
where $k'j'l'$ is one of the triples $333$, $330$, $303$, $033$,
and $kjl$ is one of the triples $122$, $212$, $221$. Now we try to
determine the exact shape of the FR. The FR is a cone given by
\begin{equation}\label{nl}
    (1\pm P_{1})^{2}=P_{2}^{2}+P_{3}^{2}
\end{equation}
(for a proof, see appendix A), where
$$
P_{_{j}}=Tr(Q_{_{j}}^{^{\mathrm{Co}}}|\alpha\rangle|\beta\rangle|\gamma\rangle\langle\alpha|\langle\beta|\langle\gamma|)
,\quad j=1,2,3.
$$
We assert that any plane tangent to the FR corresponds to an EW.
To show this, we maximize the function
\begin{equation}\label{func}
    f(P_{1},P_{2},P_{3})=A_{_{1}}P_{1}+A_{_{2}}P_{2}+A_{_{3}}P_{3}
\end{equation}
where $A_{_{i}}$ are real parameters, under the constraint
(\ref{nl}). This is a convex optimization problem since the
function and its constraint are both convex functions. Using the
Lagrange multiplier method shows that this maximum is $\pm
A_{_{1}}$ provided that $A_{_{1}}^{2}=A_{_{2}}^{2}+A_{_{3}}^{2}$.
It is easy to see that the plane
$A_{_{1}}P_{1}+A_{_{2}}P_{2}+A_{_{3}}P_{3}=\pm A_{_{1}}$ is
tangent to the surface (\ref{nl}) at the point $(-A_{_{1}}\pm 1,
A_{_{2}}, A_{_{3}})$. This plane can be rewritten as
$$
\min_{_{|\alpha\rangle|\beta\rangle|\gamma\rangle}}Tr([A_{_{1}}III\pm
A_{_{1}}Q_{_{1}}^{^{\mathrm{Co}}}\pm
(A_{_{2}}Q_{_{2}}^{^{\mathrm{Co}}}+A_{_{3}}Q_{_{3}}^{^{\mathrm{Co}}})]
|\alpha\rangle|\beta\rangle|\gamma\rangle\langle\alpha|\langle\beta|\langle\gamma|)=0.
$$
Thus the operator
$$
W_{\pm}^{^{\mathrm{Co}}}=A_{_{1}}III\pm
A_{_{1}}Q_{_{1}}^{\mathrm{Co}}\pm
(A_{_{2}}Q_{_{2}}^{\mathrm{Co}}+A_{_{3}}Q_{_{3}}^{\mathrm{Co}})
$$
has non-negative expectation value over all pure product states,
hence it can be a linear EW. By defining
$\cos\psi=\frac{A_{2}}{A_{1}}$ and $\sin\psi=\frac{A_{3}}{A_{1}}$,
$W_{\pm}^{\mathrm{Co}}$ is rewritten as
\begin{equation}\label{nlwit1}
    ^{k'j'l'}W_{kjl,i\pm}^{^{\mathrm{Co}}}= III\pm O_{k'j'l'}
    +\cos\psi\left(O_{111}+(-1)^{i}O_{kjl}\right)+\sin\psi\left(O_{lkj}+(-1)^{i}O_{jlk}\right).
\end{equation}
where $i=0,1$. Now we obtain non-linear functionals of $\rho$,
hence non-linear EWs, by optimizing
$Tr[(^{k'j'l'}W_{kjl,i\pm}^{^{\mathrm{Co}}})\rho]$ with
appropriate choice of the parameter $\psi$ as a functional of
$\rho$. We note that
$$
Tr[(^{k'j'l'}W_{kjl,i\pm}^{^{\mathrm{Co}}})\rho]=1\pm
r_{_{k'j'l'}}+\cos\psi(r_{_{111}}+(-1)^{i}r_{_{kjl}})+\sin\psi(r_{_{lkj}}+(-1)^{i}r_{_{jlk}}).
$$
By defining
$$
\cos\eta=\frac{r_{_{111}}+(-1)^{i}r_{_{kjl}}}
{\sqrt{(r_{_{111}}+(-1)^{i}r_{_{kjl}})^{2}+(r_{_{lkj}}+(-1)^{i}r_{_{jlk}})^{2}}}\,
$$
$Tr[(^{k'j'l'}W_{kjl,i\pm}^{^{\mathrm{Co}}})\rho]$ can be
rewritten as
$$
Tr[(^{k'j'l'}W_{kjl,i\pm}^{^{\mathrm{Co}}})\rho]=1\pm
r_{_{k'j'l'}}+\sqrt{(r_{_{111}}+(-1)^{i}r_{_{kjl}})^{2}+(r_{_{lkj}}
+(-1)^{i}r_{_{jlk}})^{2}}\;\cos(\psi-\eta).
$$
The trace take its minimum for $\psi-\eta=\pi$:
\begin{equation}\label{nlwit2}
    ^{k'j'l'}F_{kjl,i\pm}^{^{Co}}(\rho)=\min Tr[(^{k'j'l'}W_{kjl,i\pm}^{^{\mathrm{Co}}})\rho]=1\pm r_{_{k'j'l'}}
    -\sqrt{(r_{_{111}}+(-1)^{i}r_{_{kjl}})^{2}+(r_{_{lkj}}+(-1)^{i}r_{_{jlk}})^{2}}.
\end{equation}
These are the required non-linear functionals, hence non-linear
EWs, associated with $\rho$. It is seen that the number of such
non-linear EWs is 48.
\par
We can obtain other 48 linear EWs from
$^{k'j'l'}W_{kjl,i\pm}^{^{^{\mathrm{Co}}}}$ by conjugating them
with $MII$. This gives further 48 non-linear EWs of conical case
as follows
\begin{equation}\label{nlwit2n}
\begin{array}{c}
 \hspace{-5.6cm} ^{k'j'l'}F_{kjl,i\pm}^{'^{Co}}(\rho)=\min Tr[MII(^{k'j'l'}W_{kjl,i\pm}^{^{\mathrm{Co}}})M^{\dagger}II\rho] \\
  =1\pm r_{_{k'j'l'}}-\sqrt{(r_{_{222}}+(-1)^{i}r_{_{kjl}})^{2}+(r_{_{lkj}}+(-1)^{i}r_{_{jlk}})^{2}}\ ,\\
\end{array}
\end{equation}
here $kjl$ is one of the triples $211$, $121$ and $112$.
\par
In this way, we have constructed 96 non-linear EWs with conical
FR.
\subsection{\small{EWs with cylindrical FR}}
The second type of non-linear EWs for $\rho$ can be derived by
considering the following operators
$$
\hspace{-2cm}  Q_{_{1}}^{^{\mathrm{Cy}}}=O_{k'j'l'},\quad
  Q_{_{2}}^{^{\mathrm{Cy}}}=O_{111}+(-1)^{i}O_{kjl},\quad
Q_{_{3}}^{^{\mathrm{Cy}}}=O_{lkj}+(-1)^{i+1}O_{jlk},\quad i=0,1,\\
$$
where $k'j'l'$ is one of the triples $300$, $030$, $003$, and
$kjl$ is one of the triples $122$, $212$, $221$. It can be shown
that the FR has the cylindrical shape
\begin{equation}\label{nl2}
    P_{1}^{2}+(P_{2}+P_{3})^{2}=1,
\end{equation}
the maximum of the function (\ref{func}) under the constraint
(\ref{nl2}) is $\sqrt{A_{_{1}}^{2}+A_{_{2}}^{2}}$ provided that
$A_{_{2}}=A_{_{3}}$ and this leads to the linear EWs
\begin{equation}\label{nlwit3}
    ^{k'j'l'}W_{kjl;i_{_{1}}i_{_{2}}}^{^{\mathrm{Cy}}}=III+(\cos\psi)O_{k'j'l'}
    +\sin\psi(O_{111} +(-1)^{i_{_{1}}}O_{kjl}
+(-1)^{i_{_{2}}}O_{lkj}+(-1)^{i_{_{1}}+i_{_{2}}+1}O_{jlk}).
\end{equation}
where $\cos\psi=A_{_{1}}/\sqrt{A_{_{1}}^{2}+A_{_{2}}^{2}}\;$ and
$i_{_{1}},i_{_{2}}=0,1$.
 Similar arguments as above shows that $^{k'j'l'}W_{kjl;i_{_{1}}i_{_{2}}}^{^{\mathrm{Cy}}}$
gives rise to non-linear EWs for $\rho$ as follows
\begin{equation}\label{nlwit4}
\begin{array}{c}
  \hspace{-7cm}^{k'j'l'}F_{kjl;i_{_{1}}i_{_{2}}}^{^{Cy}}(\rho)=\min Tr[(^{k'j'l'}W_{kjl;i_{_{1}}i_{_{2}}}^{^{\mathrm{Cy}}})\rho] \\
  \hspace{2.9cm}=1-\sqrt{r_{_{k'j'l'}}^{2}+(r_{_{111}}+(-1)^{i_{_{1}}}r_{_{kjl}}
  +(-1)^{i_{_{2}}}r_{_{lkj}}+(-1)^{i_{_{1}}+i_{_{2}}+1}r_{_{jlk}})^{2}}\ . \\
\end{array}
\end{equation}
The number of these non-linear EWs is 36. We obtain other 36
non-linear EWs of this type by conjugating
$^{k'j'l'}W_{kjl,i_{_{1}}i_{_{2}}}^{^{\mathrm{Cy}}}$ with $MII$ as
follows
\begin{equation}\label{nlwit3n}
\begin{array}{c}
 \hspace{-5.6cm} ^{k'j'l'}F_{kjl;i_{_{1}}i_{_{2}}}^{'^{Cy}}(\rho)=\min Tr[MII(^{k'j'l'}W_{kjl,i_{_{1}}i_{_{2}}}^{^{\mathrm{Cy}}})M^{\dagger}II\rho]\\
  =1-\sqrt{r_{_{k'j'l'}}^{2}+(r_{_{222}}+(-1)^{i_{_{1}}}r_{_{kjl}}+(-1)^{i_{_{2}}}r_{_{lkj}}+(-1)^{i_{_{1}}+i_{_{2}}+1}r_{_{jlk}})^{2}}\ ,\\
\end{array}
\end{equation}
here $kjl$ is one of the triples $211$, $121$ and $112$ and
$i_{_{1}},i_{_{2}}=0,1$.
\par
In this way, we have constructed 72 non-linear EWs with
cylindrical FR.
\subsection{\small{EWs with spherical FR}}
The third type of non-linear EWs for $\rho$ follows from the
operators
$$
\hspace{-2cm}  Q_{_{1}}^{^{\mathrm{Sp}}}=O_{k'j'l'},\quad
  Q_{_{2}}^{^{\mathrm{Sp}}}=O_{111}+(-1)^{i}O_{kjl},\quad
Q_{_{3}}^{^{\mathrm{Sp}}}=O_{lkj}+(-1)^{i}O_{jlk},\quad i=0,1,\\
$$
where $k'j'l'$ is one of the triples $300$, $030$, $003$, and
$kjl$ is one of the triples $122$, $212$, $221$. In this case, the
FR is of spherical shape
\begin{equation}\label{nl3}
    P_{1}^{2}+P_{2}^{2}+P_{3}^{2}=1,
\end{equation}
the maximum of the function (\ref{func}) under the constraint
(\ref{nl3}) is $\sqrt{A_{_{1}}^{2}+A_{_{2}}^{2}+A_{_{3}}^{2}}$ and
this leads to the linear EW
\begin{equation}\label{nlwit5}
    ^{k'j'l'}W_{kjl,i}^{^{\mathrm{Sp}}}=III+(\sin\eta\cos\zeta)O_{k'j'l'}+\sin\eta\sin\zeta
    (O_{111}+(-1)^{i}O_{kjl})+\cos\eta
    (O_{lkj}+(-1)^{i}O_{jlk}),
\end{equation}
where
$$
\sin\eta\cos\zeta=\frac{A_{_{1}}}{\sqrt{A_{_{1}}^{2}+A_{_{2}}^{2}+A_{_{3}}^{2}}},\quad
\sin\eta\sin\zeta=\frac{A_{_{2}}}{\sqrt{A_{_{1}}^{2}+A_{_{2}}^{2}+A_{_{3}}^{2}}},\quad
\cos\eta=\frac{A_{_{3}}}{\sqrt{A_{_{1}}^{2}+A_{_{2}}^{2}+A_{_{3}}^{2}}}\
.
$$
The 18 non-linear EWs which correspond to
$^{k'j'l'}W_{kjl,i}^{^{\mathrm{Sp}}}$ is
\begin{equation}\label{nlwit6}
    ^{k'j'l'}F_{kjl;i}^{^{Sp}}(\rho)=\mathrm{min}\ Tr[(^{k'j'l'}W_{kjl,i}^{^{\mathrm{Sp}}})\rho]=
    1-\sqrt{r_{_{k'j'l'}}^{2}+(r_{_{111}}+(-1)^{i}r_{_{kjl}})^{2}+(r_{_{lkj}}
+(-1)^{i}r_{_{jlk}})^{2}}.
\end{equation}
We obtain other 18 non-linear EWs of this type by conjugating
$^3W_{i}^{Nl}$ with $MII$ as follows
\begin{equation}\label{nlwit6n}
\begin{array}{c}
 \hspace{-5.6cm} ^{k'j'l'}F_{kjl;i}^{'^{Sp}}(\rho)=\min Tr[MII(^{k'j'l'}W_{kjl,i}^{^{\mathrm{Sp}}})M^{\dagger}II\rho]\\
  =1-\sqrt{r_{_{k'j'l'}}^{2}+(r_{_{222}}+(-1)^{i}r_{_{kjl}})^{2}+(r_{_{lkj}}
+(-1)^{i}r_{_{jlk}})^{2}}.\\
\end{array}
\end{equation}
here $kjl$ is one of the triples $211$, $121$ and $112$.
\par
In this way, we have constructed 36 non-linear EWs with spherical
FR.
\section{Optimality  of the EWs}
In this section we discuss the optimality of EWs introduced so
far. Let us recall that if there exist $\epsilon> 0$ and a
positive operator $ \mathcal{P}$ such that $W'=W-\epsilon
\mathcal{P}$ be again an EW, the EW $W$ is not optimal, otherwise
it is. Every positive operator can be expressed as a sum of pure
projection operators with positive coefficients, i.e.,
$\mathcal{P}=\sum_{i}\lambda_{_{i}}|\psi_{_{i}}\rangle\langle\psi_{_{i}}|$
with all $\lambda_{_{i}}\geq0$, so we can take $\mathcal{P}$  as
pure projection operator $\mathcal{P}=|\psi\rangle\langle\psi|$.
If $W'$ is to be an  EW, then $|\psi\rangle$ must be orthogonal to
all pure product states that the expectation value of W over them
is zero. The eigenstates of each three-qubit Pauli group operator
can be chosen as pure product states, half with eigenvalue +1 and
the other half with eigenvalue -1. In EWs introduced so far, there
exists no pair of locally commuting Pauli group operators, so the
expectation value of such pauli group operators vanishes over the
pure product eigenstates of one of them.
\par
Regarding the above facts, now we are ready to discuss the
optimality of introduced EWS.
\subsection{\small{Optimality of EWs with polygonal FR}}

 Let us begin with EWs of (\ref{lew}). We
discuss two cases $i_{_{1}}=0$ and $i_{_{1}}=1$ separately. For
the case $i_{_{1}}=0$, note that as eigenstates of the operator
$\sigma_{z}\sigma_{z}\sigma_{z}$ with eigenvalue +1 we can take
the pure product states
\begin{equation}\label{eigen1}
    |z;+\rangle|z;+\rangle|z;+\rangle,\quad
|z;+\rangle|z;-\rangle|z;-\rangle,\quad
|z;-\rangle|z;+\rangle|z;-\rangle,\quad
|z;-\rangle|z;-\rangle|z;+\rangle,\quad
\end{equation}
and as eigenstates with eigenvalue -1 we can take the following
ones
\begin{equation}\label{eigen2}
    |z;+\rangle|z;+\rangle|z;-\rangle,\quad
|z;+\rangle|z;-\rangle|z;+\rangle,\quad
|z;-\rangle|z;+\rangle|z;+\rangle,\quad
|z;-\rangle|z;-\rangle|z;-\rangle.
\end{equation}
The EWs $W_{0i_{_{2}}i_{_{3}}i_{_{4}}}^{^{\mathrm{Po}}}$ have zero
expectation values over the states of (\ref{eigen2}), so if there
exists a pure projection operator $|\psi\rangle\langle\psi|$ that
can be subtracted from EWs
$W_{0i_{_{2}}i_{_{3}}i_{_{4}}}^{^{\mathrm{Po}}}$, the state
$|\psi\rangle$ ought to be of the form
\begin{equation}\label{psi1}
\begin{array}{c}
  |\psi\rangle=a_{_{+++}}|z;+\rangle|z;+\rangle|z;+\rangle+a_{_{+--}}|z;+\rangle|z;-\rangle|z;-\rangle\\
 \hspace{1.3cm} +a_{_{-+-}}|z;-\rangle|z;+\rangle|z;-\rangle+a_{_{--+}}|z;-\rangle|z;-\rangle|z;+\rangle.\\
\end{array}
\end{equation}
Expectation values of $W_{00i_{_{3}}i_{_{4}}}^{^{\mathrm{Po}}}$
over pure product eigenstates of the operator
$\sigma_{x}\sigma_{x}\sigma_{x}$ with eigenvalue -1 are zero, so
$|\psi\rangle$ should be orthogonal to these eigenstates. Applying
the orthogonality constraints gives the following equations
$$
\begin{array}{c}
  \langle x;+|\langle x;+|\langle x;-||\psi\rangle=\frac{1}{2\sqrt{2}}(a_{_{+++}}-a_{_{+--}}-a_{_{-+-}}+a_{_{--+}})=0,\\
  \langle x;+|\langle x;-|\langle x;+||\psi\rangle=\frac{1}{2\sqrt{2}}(a_{_{+++}}-a_{_{+--}}+a_{_{-+-}}-a_{_{--+}})=0,\\
  \langle x;-|\langle x;+|\langle x;+||\psi\rangle=\frac{1}{2\sqrt{2}}(a_{_{+++}}+a_{_{+--}}-a_{_{-+-}}-a_{_{--+}})=0,\\
  \langle x;-|\langle x;-|\langle x;-||\psi\rangle=\frac{1}{2\sqrt{2}}(a_{_{+++}}+a_{_{+--}}+a_{_{-+-}}+a_{_{--+}})=0.\\
\end{array}
$$
The solution of this system of four linear equations is
$a_{_{+++}}=a_{_{+--}}=a_{_{-+-}}=a_{_{--+}}=0$. Thus
$|\psi\rangle=0$, that is, there exists no pure projection
operator $|\psi\rangle\langle\psi|$, hence no positive operator
$\mathcal{P}$, which can be subtracted from
$W_{00i_{_{3}}i_{_{4}}}^{^{\mathrm{Po}}}$ and leave them EWs
again. So the EWs $W_{00i_{_{3}}i_{_{4}}}^{^{\mathrm{Po}}}$ are
optimal. Similar argument proves the optimality of EWs
$W_{01i_{_{3}}i_{_{4}}}^{^{\mathrm{Po}}}$.
\par
As for EWs $W_{1i_{_{2}}i_{_{3}}i_{_{4}}}^{^{\mathrm{Po}}}$, the
state $|\psi\rangle$ (if exises) ought to be of the form
\begin{equation}\label{psi2}
\begin{array}{c}
  |\psi\rangle=a_{_{++-}}|z;+\rangle|z;+\rangle|z;-\rangle+a_{_{+-+}}|z;+\rangle|z;-\rangle|z;+\rangle\\
 \hspace{1.3cm} +a_{_{-++}}|z;-\rangle|z;+\rangle|z;+\rangle+a_{_{---}}|z;-\rangle|z;-\rangle|z;-\rangle.\\
\end{array}
\end{equation}
The same argument as above shows the impossibility of existing
such $|\psi\rangle$. Therefore, the EWs
$W_{1i_{_{2}}i_{_{3}}i_{_{4}}}^{^{\mathrm{Po}}}$ are also optimal.
\subsection{\small{Optimality of EWs with conical FR}}
The optimality of EWs $^{330}W_{122,i\pm}^{^{\mathrm{Co}}}$ has been
proved in \cite{Ja2008}, so we talk about the optimality of EWs
$^{333}W_{122,i\pm}^{^{\mathrm{Co}}}$. Let us first find pure
product states that the expectation value of
$^{333}W_{122,i\pm}^{^{\mathrm{Co}}}$ over them vanishes. For this
purpose, we consider a pure product state as follows
\begin{equation}\label{pupr}
|\nu\rangle=\bigotimes_{j=1}^{3}\left(\cos(\frac{\theta_{_{j}}}{2})
|z;+\rangle+\exp(i\varphi_{_{j}})\sin(\frac{\theta_{_{j}}}{2})|z;-\rangle\right)
\end{equation}
and attempt to choose parameters $\theta_{_{j}}$ and
$\varphi_{_{j}}$ such that
$Tr[(^{333}W_{122,i\pm}^{^{\mathrm{Co}}})|\nu\rangle\langle\nu|]=0$.
By direct calculation, this trace is
\begin{equation}\label{trace1}
\begin{array}{c}
 \hspace{-5cm} Tr[(^{333}W_{122,i\pm}^{^{\mathrm{Co}}})|\nu\rangle\langle\nu|]
 =1\pm \cos\theta_{_{1}}\cos\theta_{_{2}}\cos\theta_{_{3}}+\sin\theta_{_{1}}\sin\theta_{_{2}}\sin\theta_{_{3}}\\
  \times[\cos\psi\cos\varphi_{_{1}}\cos(\varphi_{_{2}}+(-1)^{i+1}\varphi_{_{3}})
+\sin\psi\sin\varphi_{_{1}}\sin(\varphi_{_{2}}+(-1)^{i}\varphi_{_{3}})]. \\
\end{array}
\end{equation}
It is easy to see that the following four choices of parameters
$\theta_{_{j}}$ and $\varphi_{_{j}}$ lead to zero value for the
trace of $^{333}W_{122,0\pm}^{^{\mathrm{Co}}}$ :
$$
\begin{array}{c}
  \hspace{-.4cm} |\nu_{_{1+}}\rangle:\quad
\theta_{_{2}}=\theta_{_{3}}=\frac{\pi}{2},\quad \theta_{_{1}}
  =\frac{3\pi}{2},\quad \varphi_{_{1}}=\psi,\qquad \varphi_{_{2}}=\varphi_{_{3}}=\frac{\pi}{4},\\
  \hspace{-.4cm} |\nu_{_{2+}}\rangle:\quad
\theta_{_{1}}=\theta_{_{3}}=\frac{\pi}{2},\quad \theta_{_{2}}
  =\frac{3\pi}{2},\quad \varphi_{_{1}}=\psi,\qquad \varphi_{_{2}}=\varphi_{_{3}}=\frac{\pi}{4},\\
   |\nu_{_{3+}}\rangle:\quad \theta_{_{2}}=\theta_{_{3}}=\frac{\pi}{2},\quad \theta_{_{1}}
  =\frac{3\pi}{2},\quad \varphi_{_{1}}=-\psi,\quad \varphi_{_{2}}=\varphi_{_{3}}=-\frac{\pi}{4},\\
   |\nu_{_{4+}}\rangle:\quad \theta_{_{1}}=\theta_{_{3}}=\frac{\pi}{2},\quad \theta_{_{2}}
  =\frac{3\pi}{2},\quad \varphi_{_{1}}=-\psi,\quad \varphi_{_{2}}=\varphi_{_{3}}=-\frac{\pi}{4}.\\
\end{array}
$$
For $^{333}W_{122,0+}^{^{\mathrm{Co}}}$, the state $|\psi\rangle$
(if exists) must be of the form (\ref{psi1}) and be orthogonal to
the above four states, i.e.,
$$
\begin{array}{c}
  \langle\nu_{_{1+}}|\psi\rangle=\frac{1}{2\sqrt{2}}
  [-a_{_{+++}}+ia_{_{+--}}+\exp(-i(\psi+\frac{\pi}{4}))(a_{_{-+-}}+a_{_{--+}})]=0,\\
  \langle\nu_{_{2+}}|\psi\rangle=\frac{1}{2\sqrt{2}}
  [-a_{_{+++}}-ia_{_{+--}}-\exp(-i(\psi+\frac{\pi}{4}))(a_{_{-+-}}-a_{_{--+}})]=0,\\
  \langle\nu_{_{3+}}|\psi\rangle=\frac{1}{2\sqrt{2}}
  [-a_{_{+++}}-ia_{_{+--}}+\exp(i(\psi+\frac{\pi}{4}))(a_{_{-+-}}+a_{_{--+}})]=0,\\
  \langle\nu_{_{4+}}|\psi\rangle=\frac{1}{2\sqrt{2}}
  [-a_{_{+++}}+ia_{_{+--}}-\exp(i(\psi+\frac{\pi}{4}))(a_{_{-+-}}-a_{_{--+}})]=0.\\
\end{array}
$$
The above system of four equations has trivial solution
$a_{_{+++}}=a_{_{+--}}=a_{_{-+-}}=a_{_{--+}}=0$ provided that
$\psi\neq \pm\frac{\pi}{4},\pm\frac{3\pi}{4}$. This proves the
optimality of $^{333}W_{122,0+}^{^{\mathrm{Co}}}$ for all but
$\pm\frac{\pi}{4},\pm\frac{3\pi}{4}$ values of $\psi$. Similarly,
the optimality of $^{333}W_{122,0-}^{^{\mathrm{Co}}}$ is proved
for the same values of $\psi$.
\section{Detection of $\rho$ by EWs}
In this section, we consider the problem of detection of $\rho$ by
introduced EWs.
\subsection{\small{Detection of EWs with polygonal FR}}
First we begin with 16 EWs
$^{1}W_{i_{_{1}}i_{_{2}}i_{_{3}}i_{_{4}}}^{\mathrm{Po}}$ of
(\ref{lew}). For these EWs we have
\begin{equation}\label{det1}
   Tr( ^{1}W_{i_{_{1}}i_{_{2}}i_{_{3}}i_{_{4}}}^{^{Po}}\rho )=1+(-1)^{i_{_{1}}}r_{_{333}}
    +(-1)^{i_{_{2}}}r_{_{111}}+(-1)^{i_{_{3}}}r_{_{122}}+(-1)^{i_{_{4}}}r_{_{212}}
    +(-1)^{i_{_{2}}+i_{_{3}}+i_{_{4}}+1}r_{_{221}}.
\end{equation}
It is seen that $\rho$ is detectable by
$^{1}W_{i_{_{1}}i_{_{2}}i_{_{3}}i_{_{4}}}^{^{Po}}$ if the
parameters of $\rho$ satisfy the following conditions
\begin{equation}\label{cond1}
    b+c+\frac{1}{a}+\frac{1}{d}< \pm 4 r_{_{j}}
    \cos\varphi_{_{j}},\quad a+d+\frac{1}{b}+\frac{1}{c}< \pm 4 r_{_{j}}
    \cos\varphi_{_{j}},\quad j=1,2,3,4.
\end{equation}
\par
For the 16 EWs $^{2}W_{i_{_{1}}i_{_{2}}i_{_{3}}i_{_{4}}}^{^{Po}}$
of (\ref{lew1}), we have
\begin{equation}\label{det2}
   Tr ( ^{2}W_{i_{_{1}}i_{_{2}}i_{_{3}}i_{_{4}}}^{^{Po}} \rho ) =1+(-1)^{i_{_{1}}}r_{_{333}}
    +(-1)^{i_{_{2}}}r_{_{211}}+(-1)^{i_{_{3}}}r_{_{222}}+(-1)^{i_{_{4}}+1}r_{_{112}}
    +(-1)^{i_{_{2}}+i_{_{3}}+i_{_{4}}}r_{_{121}}.
\end{equation}
The detection condition imposes the following constraints on the
parameters
\begin{equation}\label{cond2}
    b+c+\frac{1}{a}+\frac{1}{d}< \pm 4 r_{_{j}}
    \sin\varphi_{_{j}},\quad a+d+\frac{1}{b}+\frac{1}{c}< \pm 4 r_{_{j}}
    \sin\varphi_{_{j}},\quad j=1,2,3,4.
\end{equation}
\subsection{\small{Detection of EWs with conical FR}}
The detection conditions obtained from 48 non-linear EWs
$^{k'j'l'}F_{kjl;i\pm}^{^{Co}}(\rho)$ of (\ref{nlwit2}) together
with 48 non-linear EWs $^{k'j'l'}F_{kjl;i\pm}^{^{Co}}(\rho)$ of
(\ref{nlwit2n}) are
\begin{equation}\label{cond3}
    \begin{array}{c}
       (a+\frac{1}{a}+b+\frac{1}{b})^{2}<4w\quad,\quad (a+\frac{1}{a}+c+\frac{1}{c})^{2}<4w \\
       (a+\frac{1}{a}+d+\frac{1}{d})^{2}<4w\quad,\quad (b+\frac{1}{b}+c+\frac{1}{c})^{2}<4w \\
       (b+\frac{1}{b}+d+\frac{1}{d})^{2}<4w\quad,\quad (c+\frac{1}{c}+d+\frac{1}{d})^{2}<4w \\
       (a+\frac{1}{b}+d+\frac{1}{c})^{2}<4w\quad,\quad (b+\frac{1}{a}+c+\frac{1}{d})^{2}<4w \\
    \end{array}
\end{equation}
where $w=u_{_{1}},u_{_{2}},u_{_{3}};v_{_{1}},v_{_{2}},v_{_{3}}$
and
\begin{equation}\label{uv}
    \begin{array}{c}
  u_{_{1}}=(r_{_{2}}\cos\varphi_{_{2}}\pm r_{_{3}}\cos\varphi_{_{3}})^{2}
  +(r_{_{1}}\cos\varphi_{_{1}}\mp r_{_{4}}\cos\varphi_{_{4}})^{2}, \\
   u_{_{2}}=(r_{_{1}}\cos\varphi_{_{1}}\pm r_{_{3}}\cos\varphi_{_{3}})^{2}
  +(r_{_{2}}\cos\varphi_{_{2}}\mp r_{_{4}}\cos\varphi_{_{4}})^{2}, \\
  u_{_{3}}=(r_{_{1}}\cos\varphi_{_{1}}\pm r_{_{2}}\cos\varphi_{_{2}})^{2}
  +(r_{_{3}}\cos\varphi_{_{3}}\mp r_{_{4}}\cos\varphi_{_{4}})^{2}, \\
   v_{_{1}}=(r_{_{2}}\sin\varphi_{_{2}}\pm r_{_{3}}\sin\varphi_{_{3}})^{2}
  +(r_{_{1}}\sin\varphi_{_{1}}\mp r_{_{4}}\sin\varphi_{_{4}})^{2}, \\
  v_{_{2}}=(r_{_{1}}\sin\varphi_{_{1}}\pm r_{_{3}}\sin\varphi_{_{3}})^{2}
  +(r_{_{2}}\sin\varphi_{_{2}}\mp r_{_{4}}\sin\varphi_{_{4}})^{2}, \\
   v_{_{3}}=(r_{_{1}}\sin\varphi_{_{1}}\pm r_{_{2}}\sin\varphi_{_{2}})^{2}
  +(r_{_{3}}\sin\varphi_{_{3}}\mp r_{_{4}}\sin\varphi_{_{4}})^{2}. \\
    \end{array}
\end{equation}
\subsection{\small{Detection of EWs with cylindrical FR}}
The detection conditions obtained from 36 non-linear EWs
$^{k'j'l'}F_{kjl;i_{_{1}}i_{_{2}}}^{^{Cy}}(\rho)$ of
(\ref{nlwit4}) together with 36 non-linear EWs
$^{k'j'l'}F_{kjl;i_{_{1}}i_{_{2}}}^{'^{Cy}}(\rho)$ of
(\ref{nlwit3n}) are
\begin{equation}\label{dcy}
    z_{_{i}}<16r_{_{j}}^{2}\cos^{2}\varphi_{_{j}},\quad
z_{_{i}}<16r_{_{j}}^{2}\sin^{2}\varphi_{_{j}},\quad
i=1,2,3;\;j=1,2,3,4,
\end{equation}
where
\begin{equation}\label{zdef}
\begin{array}{c}
  z_{_{1}}=(a+b+c+d)(\frac{1}{a}+\frac{1}{b}+\frac{1}{c}+\frac{1}{d}), \\
  z_{_{2}}=(a+b+\frac{1}{c}+\frac{1}{d})(c+d+\frac{1}{a}+\frac{1}{b}), \\
  z_{_{3}}=(a+c+\frac{1}{b}+\frac{1}{d})(b+d+\frac{1}{a}+\frac{1}{c}). \\
\end{array}
\end{equation}
Unfortunately, as the following argument shows, the conditions
(\ref{dcy}) are not hold for $\rho$. We can write
$$
z_{_{1}}=4+(\frac{a}{b}+\frac{b}{a})+(\frac{a}{c}+\frac{c}{a})+(\frac{a}{d}+\frac{d}{a})+
(\frac{b}{c}+\frac{c}{b})+(\frac{b}{d}+\frac{d}{b})+(\frac{c}{d}+\frac{d}{c}).
$$
The two terms of each parenthesis are inverse of each other, so
the value of each parenthesis is greater than or equal to 2 and
hence $z_{_{1}}\geq16$, while in accord to (\ref{dcy})
$z_{_{1}}<16$. Similar arguments show that
$z_{_{2}},z_{_{3}}\geq16$, but in accord to (\ref{dcy}) they are
smaller than 16.
\subsection{\small{Detection of EWs with spherical FR}}
Finally, the detection conditions obtained from 18 non-linear EWs
$^{k'j'l'}F_{kjl;i}^{^{Sp}}(\rho)$ of (\ref{nlwit6}) together with
36 non-linear EWs $^{k'j'l'}F_{kjl;i}^{'^{Sp}}(\rho)$ of
(\ref{nlwit6n}) are
$$
z_{_{i}}<4u_{_{j}}\quad,\quad z_{_{i}}<4v_{_{j}},\quad i,j=1,2,3,
$$
where $z_{_{i}}$, $u_{_{j}}$ and $v_{_{j}}$ are defined as in
(\ref{zdef}) and (\ref{uv}).
\section{Comparison with other works}
If we put $a=b=c=d=1$, $r_{_{1}}=r_{_{2}}=1$ and
$\varphi_{_{1}}=\varphi_{_{2}}=0$, the detection conditions
(\ref{cond3}) give
$$
4<4+(r_{_{3}}\cos\varphi_{_{3}}-r_{_{4}}\cos\varphi_{_{4}})^{2}.
$$
Hence, this case is detected by our EWs unless
$r_{_{3}}\cos\varphi_{_{3}}=r_{_{4}}\cos\varphi_{_{4}}$. Further
inspection shows that if in addition
$\varphi_{_{3}}=\varphi_{_{4}}=0,\pi$, then $\rho$ is separable.
So for the choice of parameters as $a=b=c=d=1$,
$r_{_{1}}=r_{_{2}}=1$, $\varphi_{_{1}}=\varphi_{_{2}}=0$ and
$\varphi_{_{3}}=\varphi_{_{4}}=0,\pi$, the $\rho$ is separable if
and only if $r_{_{3}}=r_{_{4}}$; in agreement with Ref.
\cite{pitt1}.
\par
For the case $a=1$, $0<b,c,\frac{1}{d}<1$, $r_{_{1}}=1$,
$\varphi_{_{1}}=0$ and $r_{_{2}}=r_{_{3}}=r_{_{4}}=0$, we have
$$
 Tr( ^{1}W_{1101}^{^{Po}}\rho)=\frac{2(b+c+\frac{1}{d}-3)}{2+b+c+d+\frac{1}{b}+\frac{1}{c}+\frac{1}{d}}.
$$
This trace attains its minimum value -0.3371 at
$b=c=\frac{1}{d}=0.3798$ and hence improves the result -0.1069 at
$b=c=\frac{1}{d}=0.3460$ of Ref. \cite{hyllus1}.
\par
For the case $a=1$, $0<\frac{1}{b},\frac{1}{c},d<1$, $r_{_{1}}=1$,
$\varphi_{_{1}}=0$ and $r_{_{2}}=r_{_{3}}=r_{_{4}}=0$, we have
$$
 Tr( ^{1}W_{0101}^{^{Po}}\rho)=\frac{2(\frac{1}{b}+\frac{1}{c}+d-3)}{2+b+c+d+\frac{1}{b}+\frac{1}{c}+\frac{1}{d}}.
$$
This trace attains its minimum value -0.3371 at
$\frac{1}{b}=\frac{1}{c}=d=0.3460$.

\section{$2\otimes2\otimes d$ Chessboard Density Matrices }
We generalize previous chessboard density matrices to
$2\otimes2\otimes d$ \ case and see that the PPT conditions are
valid. EW's forms remain the same with a few changes in notation.
These methods  can be applied even for higher dimensions and for
multi-qubits although the number of EW's and classification of them
increases. Using some new algebraic notation for $2\otimes2\otimes
d$  case we can write
$$
    \rho_{_{d,\alpha,\beta,\gamma}}=\sum_{j=0}^{1}
   \ (  \ \sum_{k=1}^{d} a_{_{jk}}^{^{jk}} |0jk  \rangle \langle 0jk|
     \ + \ z_{_{\overline{j}\beta}}^{^{j\alpha}} |0j\alpha  \rangle \langle 1\overline{j}\beta|
     \ + \ \overline{z}_{_{\overline{j}\beta}}^{^{j\alpha}} |1\overline{j}\beta \rangle \langle 0j\alpha|
$$
$$
     \ + \ z_{_{\overline{j}\alpha}}^{^{j\beta}} |0j\beta  \rangle \langle 1\overline{j}\alpha|
     \ + \ \overline{z}_{_{\overline{j}\alpha}}^{^{j\beta}} |1\overline{j}\alpha  \rangle \langle 0j\beta|
     \ + \ z_{_{\overline{j}\gamma}}^{^{j\gamma}} |0j\gamma  \rangle \langle 1\overline{j}\gamma|
$$
$$
     \ + \ \overline{z}_{_{\overline{j}\gamma}}^{^{j\gamma}} |1\overline{j}\gamma  \rangle \langle 0j\gamma|
     \ + \ \frac{1}{a_{_{j\alpha}}^{^{j\alpha}}} | 1j\alpha \rangle \langle 1j\alpha|
     \ + \ \frac{1}{a_{_{j\beta}}^{^{j\beta}}} | 1j\beta \rangle \langle 1j\beta|
$$
\begin{equation}\label{}
     \ + \ \frac{1}{a_{_{j\gamma}}^{^{j\gamma}}} | 1j\gamma \rangle \langle
     1j\gamma| \ )
\end{equation}

here $ \overline{j}=0 $ if $j=1$ and vice versa  and
$$  \alpha \neq \beta = 0,...,d-1    \ , \  0\leq \alpha < \beta\leq d-1 \ , \ 0\leq \gamma \leq d-1$$
$$z_{_{\overline{j}\mu}}^{^{j\nu}}=r_{_{\overline{j}\mu}}^{^{j\nu}}\exp( \ i
\varphi_{_{\overline{j}\mu}}^{^{j\nu}} \ )   \ \ ,
\overline{z}_{_{\overline{j}\mu}}^{^{j\nu}}=r_{_{\overline{j}\mu}}^{^{j\nu}}\exp(
\ - i \varphi_{_{\overline{j}\mu}}^{^{j\nu}} \ )  $$ For given
$\alpha \ , \beta$ if \ $r_{_{\overline{j}\mu}}^{^{j\nu}} \leq 1$
for every $j , \mu , \nu$ then these type density matrices have
positive partial transposes with respect to all subsystems, i.e.,
they are PPT states. All of previous witnesses classes including
polygonal, conical, cylindrical and spherical become  witnesses for
this density matrices if we replace
\\$I_{_{2}} $ \ to \ $ I_{_{d}}$ ( $d\times d$ identity matrix )
\\$\sigma_{_{x}}$ \ to \ $ \sqrt{2}\lambda_{_{\alpha\beta}}^{+} $
\\$\sigma_{_{y}}$ \ to \ $ \sqrt{2}\lambda_{_{\alpha\beta}}^{-} $
\\$\sigma_{_{z}}$ \ to \ $ E_{_{\alpha\alpha}}-E_{_{\beta\beta}}$
\\on  third partite  of each terms of  all of previous witnesses ( see appendix C
), as we do in following subsections, where  $$  \alpha \neq \beta =
0,...,d-1 \ , \ 0\leq \alpha < \beta\leq d-1.$$
\subsection{Polygonal EW's}
With the  notations as above, for polygonal case we have  $ 32
(\frac{d(d-1)}{2})$ EW's. In analogy with (\ref{lew}) the $ 16
(\frac{d(d-1)}{2})$ EW's are
\begin{equation}\label{22d poly1}
\begin{array}{c}
  ^{1}W_{i_{_{1}}i_{_{2}}i_{_{3}}i_{_{4}}}^{^{\alpha , \beta}} =I_{_{2}}I_{_{2}}I_{_{d}}+(-1)^{i_{_{1}}}\sigma_{_{z}}\sigma_{_{z}}(E_{_{\alpha\alpha}}-E_{_{\beta\beta}})+\sqrt{2}(-1)^{i_{_{2}}}\sigma_{_{x}}\sigma_{_{x}}\lambda_{_{\alpha\beta}}^{^{+}}
    \\+\sqrt{2}(-1)^{i_{_{3}}}\sigma_{_{x}}\sigma_{_{y}}\lambda_{_{\alpha\beta}}^{^{-}}+\sqrt{2}(-1)^{i_{_{4}}}\sigma_{_{y}}\sigma_{_{x}}\lambda_{_{\alpha\beta}}^{^{-}}+\sqrt{2}(-1)^{i_{_{2}}+i_{_{3}}+i_{_{4}}+1}\sigma_{_{y}}\sigma_{_{y}}\lambda_{_{\alpha\beta}}^{^{+}}\\
\end{array}
\end{equation}

where $(i_{_{1}},i_{_{2}},i_{_{3}},i_{_{4}})\in\{0,1\}^{4}$. The
remaining $ 16 (\frac{d(d-1)}{2})$ polygonal EW's can obtain by
applying the phase-shift gate locally on the first qubit. The result
is
\begin{equation}\label{22d poly2}
\begin{array}{c}
  ^{2}W_{i_{_{1}}i_{_{2}}i_{_{3}}i_{_{4}}}^{^{\alpha , \beta}} =I_{_{2}}I_{_{2}}I_{_{d}}+(-1)^{i_{_{1}}}\sigma_{_{z}}\sigma_{_{z}}(E_{_{\alpha\alpha}}-E_{_{\beta\beta}})+\sqrt{2}(-1)^{i_{_{2}}}\sigma_{_{x}}\sigma_{_{x}}\lambda_{_{\alpha\beta}}^{^{+}}
    \\+\sqrt{2}(-1)^{i_{_{3}}}\sigma_{_{x}}\sigma_{_{y}}\lambda_{_{\alpha\beta}}^{^{-}}+\sqrt{2}(-1)^{i_{_{4}}+1}\sigma_{_{y}}\sigma_{_{x}}\lambda_{_{\alpha\beta}}^{^{-}}+\sqrt{2}(-1)^{i_{_{2}}+i_{_{3}}+i_{_{4}}}\sigma_{_{y}}\sigma_{_{y}}\lambda_{_{\alpha\beta}}^{^{+}}\\
\end{array}
\end{equation}

\subsection{Conical EW's}
We can expand the relevant density matrices in terms of Pauli and
$SU(N)$ operators for convenience ( see appendix C). In the
following relations $r_{_{ijk}}$ are coefficients of relevant
operator appearing in density matrices expansions, i.e. $r_{_{ij1}}$
is the coefficient of
$\sqrt{2}\sigma_{_{i}}\sigma_{_{j}}\lambda_{_{\alpha\beta}}^{^{+}}$,
\  $r_{_{ij2}}$ is the coefficient of
$\sqrt{2}\sigma_{_{i}}\sigma_{_{j}}\lambda_{_{\alpha\beta}}^{^{-}}$,
and $r_{_{ij3}}$ is the coefficient of
$\sigma_{_{i}}\sigma_{_{j}}(E_{_{\alpha\alpha}}-E_{_{\beta\beta}})$.\\
The $ 96 (\frac{d(d-1)}{2})$ conical EW's ( in analogy with
(\ref{nlwit2}) and (\ref{nlwit2n}) ) are
\begin{equation}\label{22d coni1}
\begin{array}{c}
    ^{k'j'l'}F_{kjl,i\pm}^{^{Co}}(\rho)=\min
    Tr[(^{k'j'l'}W_{kjl,i\pm}^{^{\mathrm{Co}}})\rho]=\\
    1\pm r_{_{k'j'l'}}
    -\sqrt{(r_{_{111}}+(-1)^{i}r_{_{kjl}})^{2}+(r_{_{lkj}}+(-1)^{i}r_{_{jlk}})^{2}}.
\end{array}
\end{equation}
where $k'j'l'$ is one of the triples $333$, $330$, $303$, $033$, and
$kjl$ is one of the triples $122$, $212$, $221$.
\begin{equation}\label{22d coni2}
\begin{array}{c}
 \hspace{-5.6cm} ^{k'j'l'}F_{kjl,i\pm}^{'^{Co}}(\rho)=\min Tr[MII(^{k'j'l'}W_{kjl,i\pm}^{^{\mathrm{Co}}})M^{\dagger}II\rho] \\
  =1\pm r_{_{k'j'l'}}-\sqrt{(r_{_{222}}+(-1)^{i}r_{_{kjl}})^{2}+(r_{_{lkj}}+(-1)^{i}r_{_{jlk}})^{2}}\ ,\\
\end{array}
\end{equation}
here $kjl$ is one of the triples $211$, $121$ and $112$. Cylindrical
and spherical EW's for $2\otimes2\otimes d$ chessboard density
matrices can construct with this procedure which are in full analogy
with equations (\ref{nlwit4}) , (\ref{nlwit3n}) , (\ref{nlwit6}) ,
(\ref{nlwit6n}). As result the number of EW's are
$236(\frac{d(d-1)}{2})$

\subsection{$2\otimes2\otimes 3$ Chessboard Density Matrices : An Example}
Now let us study density matrix for $d=3$, $\alpha=0, \beta =2  \ ,
\gamma=1 $ in some details. In this case we can expand this density
matrix in terms of Pauli and Gell-Mann operators
$\Lambda_{_{1}},\ldots,\Lambda_{_{8}}$ ( see appendix D ), and all
of previous witnesses including polygonal, conical, cylindrical and
spherical are valid if we replace
\\$I_{_{2}} $ \ to \ $ I_{_{3}}$ ( $3\times 3$ identity matrix )
\\$\sigma_{_{x}}$ \ to \ $ \sqrt{2}\lambda_{_{02}}^{+} = \Lambda_{_{4}} $
\\$\sigma_{_{y}}$ \ to \ $ \sqrt{2}\lambda_{_{02}}^{-} = \Lambda_{_{5}} $
\\$\sigma_{_{z}}$ \ to \ $ E_{_{00}}-E_{_{22}}= \frac{1}{2} ( \Lambda_{_{3}}+\sqrt{3}\Lambda_{_{8}} ) $
\\on the third partite of each terms of all of previous witnesses.
For example, using above prescription,  polygonal witness in
(\ref{lew})  can be written as

\begin{equation}\label{}
\begin{array}{c}
    ^{1}W_{i_{_{1}}i_{_{2}}i_{_{3}}i_{_{4}}}^{^{\mathrm{Po}}} =I_{_{2}}I_{_{2}}I_{_{3}}+(-1)^{i_{_{1}}} \sigma_{_{z}}\sigma_{_{z}}(\frac{1}{2} ( \Lambda_{_{3}}+\sqrt{3}\Lambda_{_{8}}
    )) \
    +(-1)^{i_{_{2}}}\sigma_{_{x}}\sigma_{_{x}}\Lambda_{_{4}} \\
    +(-1)^{i_{_{3}}}\sigma_{_{x}}\sigma_{_{y}}\Lambda_{_{5}}+ (-1)^{i_{_{4}}}\sigma_{_{y}}\sigma_{_{x}} \Lambda_{_{5}} +(-1)^{i_{_{2}}+i_{_{3}}+i_{_{4}}+1}\sigma_{_{y}}\sigma_{_{y}}\Lambda_{_{4}}
\end{array}
\end{equation}
By similar substitution, all of 236 EW's can be constructed. The
detection ratio ( the ratio of entangled density matrices detected
by all our EW's to all randomly selected density matrices ), is
listed in table 2.



\section{Numerical analysis of entanglement property of $\rho$}

In this section we deal with some numerical analysis regarding
detection ability of introduced EW's for $2\otimes2\otimes2$ and
$2\otimes2\otimes3$ chessboard density matrices. Numerical
calculation is done  on random set of relevant PPT chessboard
density matrices. Those density matrices detected by EW's are
counted and then the ratio is calculated. The percent of the volume
of phase space that can be detected by introduced EWs is as listed
in the table 1.
\begin{table}[h]
\renewcommand{\arraystretch}{1}
\addtolength{\arraycolsep}{-2pt}
$$
\begin{array}{|c|c|c|c|}\hline
  \mathrm{EWs} & \mathrm{percent \ of \ detection }& \mathrm{EWs} & \mathrm{percent \ of \ detection}\\
  \hline
  \mathrm{Polygonal} & 28.3 & \mathrm{Not \ polygonal \ but \ conical} & 0.44 \\
  \mathrm{Conical} & 18.3 & \mathrm{Not \ polygonal \ but \ spherical} & 0.0275 \\
  \mathrm{Spherical} & 0.047 & \mathrm{Polygonal \ and \ spherical} & 0.0176 \\
  \mathrm{All \ EWs} & 28.62 & \mathrm{Conical \ and \ spherical} & 0.031 \\
  \hline
\end{array}
$$
\caption{\small{The percent of detection for introduced EWs. ``Not
polygonal but conical"  means  the percent of the three-qubit PPT
density matrices $\rho$ that the polygonal EWs can not detect but
conical ones can detect.}}\label{tab1}
\renewcommand{\arraystretch}{1}
\addtolength{\arraycolsep}{-3pt}
\end{table}

\begin{table}[h]
\renewcommand{\arraystretch}{1}
\addtolength{\arraycolsep}{-2pt}

$$
\begin{array}{|c|c|}\hline
  \mathrm{EWs} & \mathrm{percent \ of \ detection }\\
  \hline
  \mathrm{All \ 236 \ EW's} & \mathrm{
\overline{R}\pm \sigma = 85.45 \pm 3.336 }  \\
  \hline
\end{array}
$$
\caption{\small{The percent of detection for introduced
$2\otimes2\otimes3$ EW's. $\overline{R}$ indicates mean ratios and
$\sigma$ is standard deviation}}\label{tab2}
\renewcommand{\arraystretch}{1}
\addtolength{\arraycolsep}{-3pt}
\end{table}


\section{Conclusion}
In this paper, we have considered  a  class of three-partite PPT
Chessboard  density matrices and via an exact convex optimization
method, have constructed various linear and non-linear EWs detecting
them. The operators participating in constructing the EWs have been
chosen such that the geometrical shape of the feasible region have
been obtained exactly. The EWs have been classified according to the
geometrical shape of relevant feasible regions. When feasible region
was not a polygon, non-linear EWs were obtained. The optimality of
EWs with polygonal and conical feasible region have been shown. The
introduced EWs were all non-decomposable, since they were able to
detect PPT entangled states. Event hough,  we have  mainly discussed
these methods for $2\otimes2\otimes2$ and $2\otimes2\otimes d$
chessboard density matrices, but they are general and one can apply
them for $d_1\otimes d_2\otimes d_3$  via some minor changes in
notation and calculations. It was shown that the detection ability
of introduced EWs is often comparable with one of EWs introduced
elsewhere. In some cases, the detection ability of EWs introduced
here is better. Finally the prescription of this work is applicable
for multi-partite  PPT Chessboard  density matrices which is under
investigation.


\newpage
 \vspace{1cm}\setcounter{section}{0}
 \setcounter{equation}{0}
 \renewcommand{\theequation}{A-\roman{equation}}
  {\Large{Appendix A}}\\
{\bf Proving the inequalities}: \\
In the following proofs, we use the abbreviations
\begin{equation}\label{}
    \begin{array}{c}
       Tr(\sigma_{i}^{(1)} \ | \alpha\rangle\langle\alpha|)=a_{_{i}} \\
       Tr(\sigma_{i}^{(2)} \ | \beta\rangle\langle\beta|)=b_{_{i}}  \\
       Tr(\sigma_{i}^{(3)} \ | \gamma\rangle\langle\gamma|)=c_{_{i}}.\\
     \end{array}
\end{equation}
Since $a_{_{1}}^{2}+a_{_{2}}^{2}+a_{_{3}}^{2}=1$ and also the
similar relations hold for $b_{_{i}}$'s and $c_{_{i}}$'s, so the
points $a,b,c$ lie on a unit sphere and we can parameterize their
coordinates by using spherical coordinates $\theta$ and $\varphi$
as follows
$$
\begin{array}{c}
  a_{_{1}}=\sin{\theta_{_{1}}}\cos{\varphi_{_{1}}},  \quad
  a_{_{2}}=\sin{\theta_{_{1}}}\sin{\varphi_{_{1}}},\quad a_{_{3}}=\cos{\theta_{_{1}}} \\
   b_{_{1}}=\sin{\theta_{_{2}}}\cos{\varphi_{_{2}}},  \quad
  b_{_{2}}=\sin{\theta_{_{2}}}\sin{\varphi_{_{2}}},\quad b_{_{3}}=\cos{\theta_{_{2}}} \\
   c_{_{1}}=\sin{\theta_{_{3}}}\cos{\varphi_{_{3}}},  \quad
  c_{_{2}}=\sin{\theta_{_{3}}}\sin{\varphi_{_{3}}},\quad c_{_{3}}=\cos{\theta_{_{3}}}. \\
\end{array}
$$
{\bf The proof of (\ref{hp})}:\\
To prove this equality, we note that
$$
P_{1}=a_{_{3}}b_{_{3}}c_{_{3}}=\cos\theta_{_{1}}\cos\theta_{_{2}}\cos\theta_{_{3}}
$$
$$
P_{2}=a_{_{1}}(b_{_{1}}c_{_{1}}\pm b_{_{2}}c_{_{2}})=
\sin\theta_{_{1}}\sin\theta_{_{2}}\sin\theta_{_{3}}\cos\varphi_{_{1}}\cos(\varphi_{_{3}}\mp\varphi_{_{2}})
$$
$$
P_{3}=a_{_{2}}(b_{_{1}}c_{_{2}}\mp b_{_{2}}c_{_{1}})=
\sin\theta_{_{1}}\sin\theta_{_{2}}\sin\theta_{_{3}}\sin\varphi_{_{1}}\sin(\varphi_{_{3}}\mp\varphi_{_{2}})
$$
whence
$$
\frac{P_{2}^{2}}{\cos^{2}\varphi_{_{1}}}+\frac{P_{3}^{2}}{\sin^{2}\varphi_{_{1}}}
=\sin^{2}\theta_{_{1}}\sin^{2}\theta_{_{2}}\sin^{2}\theta_{_{3}}
$$
Taking derivative  with respect to $\varphi_{_{1}}$ we obtain
$$
\frac{P_{2}}{\cos^{2}\varphi_{_{1}}}=\pm\frac{P_{3}}{\sin^{2}\varphi_{_{1}}}
$$
Above two equations yield
$$
\sin^{2}\varphi_{_{1}}=\frac{\pm P_{3}(P_{2}\pm
P_{3})}{\sin^{2}\theta_{_{1}}\sin^{2}\theta_{_{2}}\sin^{2}\theta_{_{3}}}
$$
$$
\cos^{2}\varphi_{_{1}}=\frac{P_{2}(P_{2}\pm
P_{3})}{\sin^{2}\theta_{_{1}}\sin^{2}\theta_{_{2}}\sin^{2}\theta_{_{3}}}
$$
Noting that $\sin^{2}\varphi_{_{1}}+\cos^{2}\varphi_{_{1}}=1$, we
get
$$
P_{2}+P_{3}=\pm\sin\theta_{_{1}}\sin\theta_{_{2}}\sin\theta_{_{3}}\quad,\quad
P_{2}-P_{3}=\pm\sin\theta_{_{1}}\sin\theta_{_{2}}\sin\theta_{_{3}}
$$
Eliminating $\theta_{_{1}}$ between $P_{2}\pm P_{3}$ and $P_{1}$
leads to
$$
\frac{(P_{2}\pm
P_{3})^{2}}{\sin^{2}\theta_{_{2}}\sin^{2}\theta_{_{3}}}+\frac{P_{1}^{2}}{\cos^{2}\theta_{_{2}}\cos^{2}\theta_{_{3}}}=1
$$
Taking derivative  with respect to $\theta_{_{2}}$ and by similar
argument as above, we get
$$
\frac{P_{1}}{\cos\theta_{_{3}}}+\frac{P_{2}\pm
P_{3}}{\sin\theta_{_{3}}}=\pm1\quad,\quad
\frac{P_{1}}{\cos\theta_{_{3}}}-\frac{P_{2}\pm
P_{3}}{\sin\theta_{_{3}}}=\pm1.
$$
Finally, taking derivative with respect to $\theta_{_{3}}$ and using
the identity $\sin^{2}\theta_{_{3}}+\cos^{2}\theta_{_{3}}=1$ gives
$$
P_{1}^{\frac{2}{3}}+(P_{2}\pm P_{3})^{\frac{2}{3}}=1
$$
But, as the Fig. 1 shows, this is a concave curve. Since the mixed
separable states are convex combinations of pure product states,
the boundaries of FR are the planes of (\ref{hp}).\\
{\bf The proof of (\ref{nl})}:\\
The proofs are similar, so we give the proof for the case
$Q_{1}^{_{Co}}=O_{333}$ and $kjl=122$. We note that
$$
P_{1}=a_{_{3}}b_{_{3}}c_{_{3}}=\cos\theta_{_{1}}\cos\theta_{_{2}}\cos\theta_{_{3}},
$$
$$
P_{2}=a_{_{1}}(b_{_{1}}c_{_{1}}\pm b_{_{2}}c_{_{2}})=
\sin\theta_{_{1}}\sin\theta_{_{2}}\sin\theta_{_{3}}\cos\varphi_{_{1}}\cos(\varphi_{_{2}}\mp\varphi_{_{3}}),
$$
$$
P_{3}=a_{_{2}}(b_{_{1}}c_{_{2}}\pm b_{_{2}}c_{_{1}})=
\sin\theta_{_{1}}\sin\theta_{_{2}}\sin\theta_{_{3}}\sin\varphi_{_{1}}\sin(\varphi_{_{2}}\pm\varphi_{_{3}}).
$$
By eliminating $\theta_{_{1}}$ and $\varphi_{_{1}}$, we get
$$
\frac{P_{1}^{2}}{\cos^{2}\theta_{_{2}}\cos^{2}\theta_{_{3}}}+\frac{1}{\sin^{2}\theta_{_{2}}\sin^{2}\theta_{_{3}}}
\left(\frac{P_{2}^{2}}{\cos^{2}(\varphi_{_{2}}\mp\varphi_{_{3}})}+\frac{P_{3}^{2}}{\sin^{2}
(\varphi_{_{2}}\pm\varphi_{_{3}})}\right)=1
$$
Now we put $\varphi_{_{2}}=\varphi_{_{3}}=\frac{\pi}{4}$ or
$\varphi_{_{2}}=\frac{3\pi}{4}$ and $\varphi_{_{3}}=\frac{\pi}{4}$
to obtain
$$
\frac{P_{1}^{2}}{\cos^{2}\theta_{_{2}}\cos^{2}\theta_{_{3}}}
+\frac{P_{2}^{2}+P_{3}^{2}}{\sin^{2}\theta_{_{2}}\sin^{2}\theta_{_{3}}}=1
$$
Derivation with respect to $\theta_{_{2}}$ leads to
$$
\frac{P_{1}}{\cos^{2}\theta_{_{2}}\cos\theta_{_{3}}}
=\pm\frac{(P_{2}^{2}+P_{3}^{2})^{\frac{1}{2}}}{\sin^{2}\theta_{_{2}}\sin\theta_{_{3}}}
$$
Above two equations yield
$$
\sin^{2}\theta_{_{2}}=\frac{(P_{2}^{2}+P_{3}^{2})^{\frac{1}{2}}}{\sin\theta_{_{3}}}\left(\pm
\frac{P_{1}}{\cos\theta_{_{3}}}+\frac{(P_{2}^{2}+P_{3}^{2})^{\frac{1}{2}}}{\sin\theta_{_{3}}}\right),\quad
\cos^{2}\theta_{_{2}}=\frac{P_{1}}{\cos\theta_{_{3}}}\left(\frac{P_{1}}{\cos\theta_{_{3}}}
\pm\frac{(P_{2}^{2}+P_{3}^{2})^{\frac{1}{2}}}{\sin\theta_{_{3}}}\right)
$$
From $\sin^{2}\theta_{_{2}}+\cos^{2}\theta_{_{2}}=1$, we have
$$
\frac{P_{1}}{\cos\theta_{_{3}}}+\frac{(P_{2}^{2}+P_{3}^{2})^{\frac{1}{2}}}{\sin\theta_{_{3}}}=\pm1,
\quad\frac{P_{1}}{\cos\theta_{_{3}}}-\frac{(P_{2}^{2}+P_{3}^{2})^{\frac{1}{2}}}{\sin\theta_{_{3}}}=\pm1.
$$
Finally, taking derivative with respect to $\theta_{_{3}}$ and using
the identity $\sin^{2}\theta_{_{3}}+\cos^{2}\theta_{_{3}}=1$ gives
$$
P_{1}^{\frac{2}{3}}+((P_{2}^{2}+P_{3}^{2})^{\frac{1}{2}})^{\frac{2}{3}}=1
$$
But, as the Fig. 1 shows, this is a concave curve in terms of
variables $P_{1}$ and $(P_{2}^{2}+P_{3}^{2})^{\frac{1}{2}}$. Since
the mixed separable states are convex combinations of pure product
states, the relations between these two variables  are given by
the lines
$$
P_{1}+(P_{2}^{2}+P_{3}^{2})^{\frac{1}{2}}=\pm1,\quad
P_{1}-(P_{2}^{2}+P_{3}^{2})^{\frac{1}{2}}=\pm1
$$
So the relations between $P_{1}$, $P_{2}$, and $P_{3}$ are as in
(\ref{nl}).
\par
If we take $Q_{1}^{^{Co}}=O_{330}$, the proof of (\ref{nl})
proceeds as follows. We note that in this case
$$
P_{1}=a_{_{3}}b_{_{3}}=\cos\theta_{_{1}}\cos\theta_{_{2}}
$$
By eliminating $\theta_{_{1}}$ and $\varphi_{_{1}}$, we get
$$
\frac{P_{1}^{2}}{\cos^{2}\theta_{_{2}}}+\frac{1}{\sin^{2}\theta_{_{2}}\sin^{2}\theta_{_{3}}}
\left(\frac{P_{2}^{2}}{\cos^{2}(\varphi_{_{2}}\mp\varphi_{_{3}})}+\frac{P_{3}^{2}}{\sin^{2}
(\varphi_{_{2}}\pm\varphi_{_{3}})}\right)=1
$$
Now we put $\theta_{_{3}}=\frac{\pi}{2}$ and
$\varphi_{_{2}}=\varphi_{_{3}}=\frac{\pi}{4}$ or
$\varphi_{_{2}}=\frac{3\pi}{4}$ and $\varphi_{_{3}}=\frac{\pi}{4}$
to obtain
$$
\frac{P_{1}^{2}}{\cos^{2}\theta_{_{2}}}+\frac{P_{2}^{2}+P_{3}^{2}}{\sin^{2}\theta_{_{2}}}=1
$$
Taking derivative with respect to $\theta_{_{2}}$ leads to
$$
\frac{P_{1}}{\cos^{2}\theta_{_{2}}}=\pm\frac{(P_{2}^{2}+P_{3}^{2})^{\frac{1}{2}}}{\sin^{2}\theta_{_{2}}}
$$
Above two equations yield
$$
\sin^{2}\theta_{_{2}}=\pm(P_{2}^{2}+P_{3}^{2})^{\frac{1}{2}}(P_{1}\pm(P_{2}^{2}+P_{3}^{2})^{\frac{1}{2}}),\quad
\cos^{2}\theta_{_{2}}=P_{1}(P_{1}\pm(P_{2}^{2}+P_{3}^{2})^{\frac{1}{2}})
$$
Finally, the (\ref{nl}) follows from the identity
$\sin^{2}\theta_{_{2}}+\cos^{2}\theta_{_{2}}=1$.\\\\

\newpage
{\Large{Appendix B}}\\
{\bf The coefficients of Pauli operators appearing in $\rho$}:\\
$$
\begin{array}{c|c}
  r_{_{111}}=\frac{2}{n}(r_{_1}\cos\varphi_{_{1}}+r_{_2}\cos\varphi_{_{2}}+r_{_3}\cos\varphi_{_{3}}+r_{_4}\cos\varphi_{_{4}}) & r_{_{300}}=\frac{1}{n}(a+b+c+d-\frac{1}{a}-\frac{1}{b}-\frac{1}{c}-\frac{1}{d}) \\
  r_{_{112}}=\frac{2}{n}(r_{_1}\sin\varphi_{_{1}}-r_{_2}\sin\varphi_{_{2}}+r_{_3}\sin\varphi_{_{3}}-r_{_4}\sin\varphi_{_{4}}) & r_{_{030}}=\frac{1}{n}(a+b-c-d-\frac{1}{a}-\frac{1}{b}+\frac{1}{c}+\frac{1}{d}) \\
  r_{_{121}}=\frac{2}{n}(r_{_1}\sin\varphi_{_{1}}+r_{_2}\sin\varphi_{_{2}}-r_{_3}\sin\varphi_{_{3}}-r_{_4}\sin\varphi_{_{4}}) & r_{_{003}}=\frac{1}{n}(a-b+c-d-\frac{1}{a}+\frac{1}{b}-\frac{1}{c}+\frac{1}{d}) \\
  r_{_{211}}=\frac{2}{n}(-r_{_1}\sin\varphi_{_{1}}-r_{_2}\sin\varphi_{_{2}}-r_{_3}\sin\varphi_{_{3}}-r_{_4}\sin\varphi_{_{4}}) & r_{_{330}}=\frac{1}{n}(a+b-c-d+\frac{1}{a}+\frac{1}{b}-\frac{1}{c}-\frac{1}{d}) \\
  r_{_{122}}=\frac{2}{n}(-r_{_1}\cos\varphi_{_{1}}+r_{_2}\cos\varphi_{_{2}}+r_{_3}\cos\varphi_{_{3}}-r_{_4}\cos\varphi_{_{4}}) & r_{_{303}}=\frac{1}{n}(a-b+c-d+\frac{1}{a}-\frac{1}{b}+\frac{1}{c}-\frac{1}{d}) \\
  r_{_{212}}=\frac{2}{n}(r_{_1}\cos\varphi_{_{1}}-r_{_2}\cos\varphi_{_{2}}+r_{_3}\cos\varphi_{_{3}}-r_{_4}\cos\varphi_{_{4}}) & r_{_{033}}=\frac{1}{n}(a-b-c+d+\frac{1}{a}-\frac{1}{b}-\frac{1}{c}+\frac{1}{d}) \\
  r_{_{221}}=\frac{2}{n}(r_{_1}\cos\varphi_{_{1}}+r_{_2}\cos\varphi_{_{2}}-r_{_3}\cos\varphi_{_{3}}-r_{_4}\cos\varphi_{_{4}}) & r_{_{333}}=\frac{1}{n}(a-b-c+d-\frac{1}{a}+\frac{1}{b}+\frac{1}{c}-\frac{1}{d}) \\
  r_{_{222}}=\frac{2}{n}(r_{_1}\sin\varphi_{_{1}}-r_{_2}\sin\varphi_{_{2}}-r_{_3}\sin\varphi_{_{3}}+r_{_4}\sin\varphi_{_{4}}) & \ \\
\end{array}
$$


\newpage
{\Large{Appendix C}}\\
{\bf  \\}Every d-dimensional square matrix could be written in terms
of square matrices $E_{_{ij}}$, which show the value 1 at the
position $(i , j )$ and zeros elsewhere. Now one can define
Hermitian traceless basis for d-dimensional matrices as follows (
see \cite{Pf} )
\\The off-diagonal basis are
$$\lambda_{\alpha \beta}^{^{+}}=\frac{1}{\sqrt{2}}( E_{_{\alpha
\beta}}+E_{_{\beta\alpha}} )$$  $$\lambda_{\alpha
\beta}^{^{-}}=\frac{1}{i \sqrt{2}}( E_{_{\alpha
\beta}}-E_{_{\beta\alpha}} )$$ \\and diagonal basis are \\ \\
 $\lambda_{_{0}}=\tiny\left(
                                                 \begin{array}{ccccc}
                                                   1 &  &  &  & 0 \\
                                                    & -1 &  &  &  \\
                                                    &  & 0 &  &  \\
                                                    &  &  & . &  \\
                                                   0 &  &  &  & 0 \\
                                                 \end{array}
                                               \right)
$ , $\lambda_{_{1}}=\tiny \frac{1}{\sqrt{3}}\left(
                                                 \begin{array}{ccccc}
                                                   1 &  &  &  & 0 \\
                                                    & 1 &  &  &  \\
                                                    &  & -2 &  &  \\
                                                    &  &  & . &  \\
                                                   0 &  &  &  & 0 \\
                                                 \end{array}
                                               \right)
$ , ... , $\lambda_{_{d-2}}=\tiny \sqrt{\frac{2}{d (d-1)}}\left(
                                                 \begin{array}{ccccc}
                                                   1 &  &  &  & 0 \\
                                                    & 1 &  &  &  \\
                                                    &  & . &  &  \\
                                                    &  &  & 1 &  \\
                                                   0 &  &  &  & -d+1 \\
                                                 \end{array}
                                               \right)
$. \\ \\In order to generalize the witnesses,  we must write
$E_{_{\alpha\alpha}}$ in terms of $I_{_{d}}$ ( $d\times d$ identity
matrix ) and $\lambda_{_{\alpha}}$'s. Some calculation shows that
$$E_{_{ii}}=E_{_{i+1,i+1}}+\sqrt{\frac{i+2}{2(i+1)}}\ \lambda_{_{i}}-\sqrt{\frac{i}{2(i+1)}}\ \lambda_{_{i-1}} \ , \ \ 0\leq i \leq d-2$$
( recursion relation ) and
\\$$E_{_{d-1,d-1}}=\frac{1}{d} I_{_{d}} - \sqrt{\frac{d-1}{2 d}}\
\lambda_{_{d-2}}$$\\

{\bf Proving the inequalities for $2\otimes2\otimes d$} \ : \\
 The proof is almost the as explained in appendix A. We use the
abbreviations
\begin{equation}\label{}
    \begin{array}{c}
       Tr(\sigma_{i}^{(1)} \ | \alpha\rangle\langle\alpha|)=a_{_{i}} \\
       Tr(\sigma_{i}^{(2)} \ | \beta\rangle\langle\beta|)=b_{_{i}}  \\
       Tr(\sqrt{2}\lambda_{_{\alpha\beta}}^{^{+}} \ | \xi\rangle\langle\xi|)=c_{_{1}}\\
       Tr(\sqrt{2}\lambda_{_{\alpha\beta}}^{^{-}} \ | \xi\rangle\langle\xi|)=c_{_{2}}\\
       Tr((E_{_{\alpha\alpha}}-E_{_{\beta\beta}}) \ | \xi\rangle\langle\xi|)=c_{_{3}}\\
     \end{array}
\end{equation}
where $$| \xi\rangle\ =
\frac{1}{\sqrt{r_{_{0}}^{2}+...+r_{_{d-1}}^{2}}}\left(
                                               \tiny \begin{array}{c}
                                                  r_{_{0}}e^{^{i \theta_{_{0}}}} \\
                                                  \vdots \\
                                                  r_{_{d-1}}e^{^{i \theta_{_{d-1}}}} \\
                                                \end{array}
                                              \right)
$$

We have $$a_{_{1}}^{2}+a_{_{2}}^{2}+a_{_{3}}^{2}=1 \ , \
b_{_{1}}^{2}+b_{_{2}}^{2}+b_{_{3}}^{2}=1$$ and
$$c_{_{1}}^{2}+c_{_{2}}^{2}+c_{_{3}}^{2}=\frac{(r_{_{\alpha}}^{2}+r_{_{\beta}}^{2})^{2}}{(r_{_{0}}^{2}+\ldots+r_{_{d-1}}^{2})^{2}}= q $$
if we set $q=1$ without  loss of generality, then the points $a,b,c$
lie on a unit sphere and we can parameterize their coordinates by
using spherical coordinates $\theta$ and $\varphi$ as follows
$$
\begin{array}{c}
  a_{_{1}}=\sin{\theta_{_{1}}}\cos{\varphi_{_{1}}},  \quad
  a_{_{2}}=\sin{\theta_{_{1}}}\sin{\varphi_{_{1}}},\quad a_{_{3}}=\cos{\theta_{_{1}}} \\
   b_{_{1}}=\sin{\theta_{_{2}}}\cos{\varphi_{_{2}}},  \quad
  b_{_{2}}=\sin{\theta_{_{2}}}\sin{\varphi_{_{2}}},\quad b_{_{3}}=\cos{\theta_{_{2}}} \\
   c_{_{1}}=\sin{\theta_{_{3}}}\cos{\varphi_{_{3}}},  \quad
  c_{_{2}}=\sin{\theta_{_{3}}}\sin{\varphi_{_{3}}},\quad c_{_{3}}=\cos{\theta_{_{3}}}. \\
\end{array}
$$


\newpage
{\Large{Appendix D}}\\
{\bf The Gell-Mann Matrices \\}

The analog of the Pauli matrices for SU(3) are Gell-Mann matrices
defined as:
\\
\\
$\Lambda_{_{1}}=\tiny\left(
                                  \begin{array}{ccc}
                                    0 & 1 & 0 \\
                                    1 & 0 & 0 \\
                                    0 & 0 & 0 \\
                                  \end{array}
                                \right)
$ \ , \ $\Lambda_{_{2}}=\tiny\left(
                                  \begin{array}{ccc}
                                    0 & -i & 0 \\
                                    i & 0 & 0 \\
                                    0 & 0 & 0 \\
                                  \end{array}
                                \right)
$ \ , \ $\Lambda_{_{3}}=\tiny\left(
                                  \begin{array}{ccc}
                                    1 & 0 & 0 \\
                                    0 & -1 & 0 \\
                                    0 & 0 & 0 \\
                                  \end{array}
                                \right)
$ \\ \\ $\Lambda_{_{4}}=\tiny\left(
                                  \begin{array}{ccc}
                                    0 & 0 & 1 \\
                                    0 & 0 & 0 \\
                                    1 & 0 & 0 \\
                                  \end{array}
                                \right)
$ \ , \ $\Lambda_{_{5}}=\tiny\left(
                                  \begin{array}{ccc}
                                    0 & 0 & -i \\
                                    0 & 0 & 0 \\
                                    i & 0 & 0 \\
                                  \end{array}
                                \right)
$ \ , \ $\Lambda_{_{6}}=\tiny\left(
                                  \begin{array}{ccc}
                                    0 & 0 & 0 \\
                                    0 & 0 & 1 \\
                                    0 & 1 & 0 \\
                                  \end{array}
                                \right)
$ \\ \\ $\Lambda_{_{7}}=\tiny\left(
                                  \begin{array}{ccc}
                                    0 & 0 & 0 \\
                                    0 & 0 & -i \\
                                    0 & i & 0 \\
                                  \end{array}
                                \right)
$ \ , \ $\Lambda_{_{8}}=\frac{1}{\sqrt{3}}\tiny\left(
                                  \begin{array}{ccc}
                                    1 & 0 & 0 \\
                                    0 & 1 & 0 \\
                                    0 & 0 & -2 \\
                                  \end{array}
                                \right)
$


\newpage
{\bf Figure Captions}

 {\bf Figure-1:} The boundaries of feasible region for pure
 product states (dotted curve) and mixed separable states (line)
 for EWs of relation (\ref{lew}).

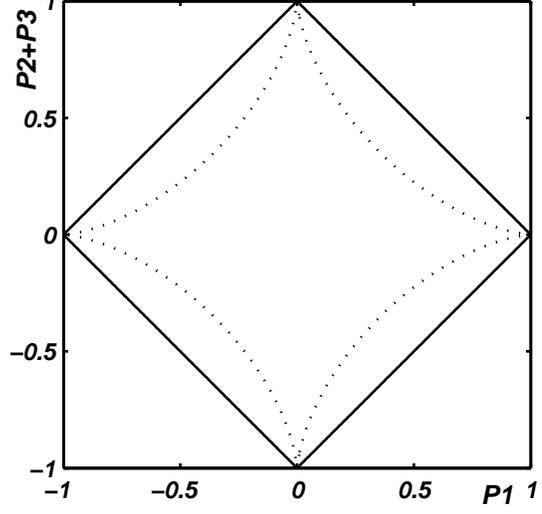
\end{document}